\documentclass[conference]{IEEEtran}
\IEEEoverridecommandlockouts
\pdfoutput=1
\usepackage[noadjust]{cite}
\usepackage[mathscr]{euscript}
\usepackage{array}
\usepackage{mdwmath}
\usepackage{mathtools}
\usepackage{subfig}
\usepackage{graphicx}
\usepackage{url}
\usepackage{tikz}
\usepackage{pgfplots}
\graphicspath{{../pdf/}{../jpeg/}}
\DeclareGraphicsExtensions{.pdf,.jpeg,.png}
\usepackage{epstopdf}
\usepackage{amsfonts}
\usepackage{booktabs}
\usepackage{lipsum}

\newcommand\blfootnote[1]{%
  \begingroup
  \renewcommand\thefootnote{}\footnote{#1}%
  \addtocounter{footnote}{-1}%
  \endgroup
}
\usepackage[shrink=50]{microtype}
\usepackage{cite}
\usepackage{amsmath,amssymb,amsfonts}
\newtheorem{theorem}{Theorem}

\usepackage{latexsym}
\usepackage{amsfonts,amssymb,amsmath}

\begin{document}

\title{\huge Channel Estimation for Distributed Intelligent Reflecting Surfaces Assisted Multi-User MISO Systems}

\author{Hibatallah Alwazani, \IEEEmembership{Student Member, IEEE}, Qurrat-Ul-Ain Nadeem, \IEEEmembership{Member, IEEE},\\  Anas Chaaban, \IEEEmembership{Senior Member, IEEE}}

\maketitle

\begin{abstract}
Intelligent reflecting surfaces (IRSs)-assisted wireless communication promises improved system performance, while posing new challenges in channel estimation (CE) due to the passive nature of the reflecting elements. Although a few CE protocols for IRS-assisted multiple-input single-output (MISO) systems have appeared, they either require long channel training times or are developed under channel sparsity assumptions. Moreover, existing works focus on a single IRS, whereas in practice multiple such surfaces should be installed to truly benefit from the concept of reconfiguring propagation environments. In light of these challenges, this paper tackles the CE problem for the distributed IRSs-assisted multi-user MISO system. An optimal CE protocol  requiring relatively low training overhead is developed using Bayesian techniques under the practical assumption that the BS-IRSs channels are dominated by the line-of-sight (LoS) components. An optimal solution for the phase shifts vectors required at all IRSs during CE is determined and the minimum mean square error (MMSE) estimates of the BS-users direct channels and the IRSs-users channels are derived. Simulation results corroborate the normalized MSE (NMSE) analysis and establish the advantage of the proposed protocol as compared to benchmark scheme in terms of training overhead.
 \end{abstract}



\section{Introduction}
\label{intro}


An intelligent reflecting surface (IRS) consists of a large number of low-cost passive reflecting elements, where each element can change the phase shift of the incident waves from the base station (BS) in a passive way to achieve reflect beamforming gains. This has made IRS a viable solution to make the cellular systems consume less overall power while promising better performance \cite{LIS_los, maxmin, renzo2020}.  Given the IRS phase shifts have to be updated in each coherence interval, channel state information (CSI) needs to be acquired for all IRS-assisted links at the pace of channel variation, which overburdens the system.
\blfootnote{\thanks{This publication is based upon work supported by the King Abdullah University of Science and Technology (KAUST) under Award No. OSR-2018-CRG7-3734}}

The passivity of the IRS translates to an inability to estimate the channels or assist the BS in estimating the channels by transmitting pilot symbols, which poses the biggest challenge in developing CE protocols for these systems. The first estimation protocol that appeared for a single-user IRS-assisted multiple-input single-output (MISO) system, known as the ON/OFF protocol \cite{mishra}, estimates all IRS-assisted links one-by-one by serially turning one element ON while keeping the others OFF. This protocol was significantly improved in terms of the normalized mean squared error (NMSE) performance in \cite{MVUE} for a single-user system and in \cite{nadeem2020intelligent} for a multi-user system, where an optimal solution for IRS phase shifts vectors in the CE phase was developed. However, the CE time scales linearly with the number of IRS elements in all these works, imposing prohibitively high training overhead.  To combat this, \cite{ofdm_irs} and \cite{elements_grouping} introduce the idea of grouping adjacent IRS elements into sub-surfaces, which will decrease the training overhead but also reduce the reflect beamforming gains promised by using a large number of IRS elements. Other solutions exploit channel sparsity that exists in mmWave and massive MIMO channels to develop low-overhead algorithms for CE \cite{los, cas}.  A two-timescale CE framework  is proposed in \cite{twotimescale} that exploits the property that the BS-IRS channel is quasi-static, while the IRS-user channel is mobile. The assumption that BS-IRS channel is LoS and fixed is also made in many other works such as  \cite{los_channel,nadeem2020intelligent,los, maxmin, LIS_los} and is quite practical given BS and IRS have fixed positions with few obstacles around.

 The existing few works on CE all deal with single IRS-assisted MISO systems and still impose large training overheads. Recently, the idea of deploying multiple IRSs in a distributed manner has been considered to overcome several signal blockages by coating the blocking structures with IRSs. Moreover, since BS-IRS channel is expected to be of low-rank in many practical cases \cite{maxmin}, the use of distributed IRSs can yield a high-rank BS-IRSs channel matrix. 


 In this paper, we tackle the CE problem for a distributed IRSs-assisted MISO system. We consider correlated Rayleigh fading channels between the BS and users, and the IRSs and users, while each BS-IRS channel is LoS. The LoS BS-IRSs channel matrices can be learnt apriori at the BS using the locations of the IRS, which are fixed and can be used to compute the LoS angle of departure (AoD) and angle of arrival (AoA). Leveraging this, we develop an optimal CE protocol using the Bayesian technique of minimum mean squared error (MMSE) estimation, that promises low CE error and imposes a significantly lower training overhead than what we need when we extend the conventional protocols from \cite{mishra, MVUE, nadeem2020intelligent} to this setup. The normalized mean squared error (NMSE) in the MMSE estimates is analytically derived and studied using simulations.
  To the best of our knowledge, this paper marks the first contribution in developing a  CE scheme for the distributed IRSs case. 
 


\section{System Model}
\begin{figure}
    \centering
    \includegraphics[scale=0.35]{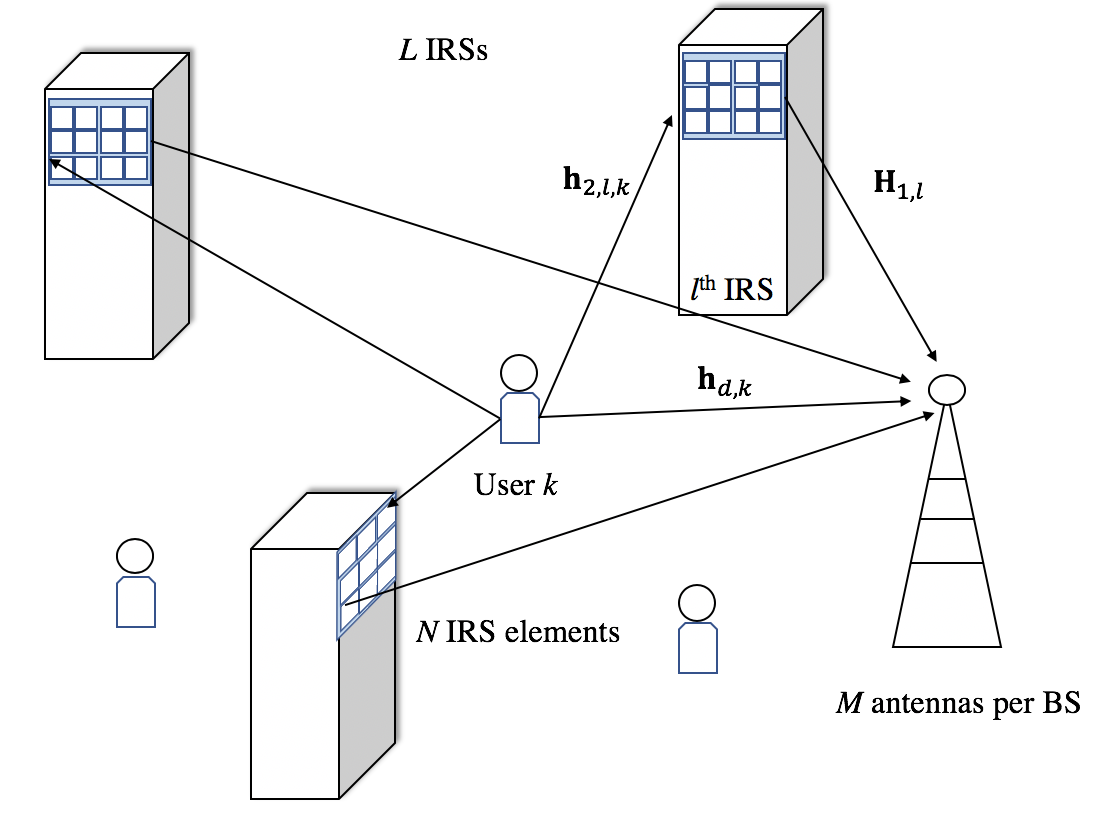}
    \caption{Distributed IRS-assisted MISO transmission model.}
    \label{Sysmodel}
\end{figure}

Consider  $M$ antennas at the BS serving $K$ single-antenna users. The communication between the BS and users is assisted by $L$ IRSs, each equipped with $N$ elements (see Fig. \ref{Sysmodel}). The IRSs are deployed in the environment in a distributed manner with fixed positions and their operation is controlled by IRS controllers that communicate with the BS over a backhaul link \cite{maxmin}.  The channel between the BS and user $k$ is given as
\begin{align}
&\mathbf{h}_{{IRS}_{k}}=\sum_{l=1}^L \mathbf{H}_{1,l}\boldsymbol{\Theta}_l\mathbf{h}_{2,l,k}+\mathbf{h}_{d,k},
\end{align}
where $\mathbf{H}_{1,l} \in \mathbb{C}^{M\times N}$ is the channel between BS and IRS $l$, $\mathbf{h}_{2,l,k}\in \mathbb{C}^{N\times 1}$ is the channel between IRS $l$ and user $k$, and $\boldsymbol{\Theta}_l=\text{diag}(\alpha_{l,1} \exp(j \theta_{l,1}), \dots, \alpha_{l,N}\exp(j\theta_{l,N}))\in \mathbb{C}^{N\times N}$ is the reflection matrix for IRS $l$, where $\theta_{l,n}\in [0,2\pi]$ is the phase-shift applied by element $n$ of IRS $l$ and $\alpha_{n,l}\in [0,1]$ is the amplitude reflection  coefficient. Also $\mathbf{h}_{d,k}\in \mathbb{C}^{N\times 1}$ is the direct channel between BS and user $k$.



Since the antennas at the BS as well as the reflecting elements in the IRSs are arranged compactly, we model the channels $\mathbf{h}_{2,l,k}$ and $\mathbf{h}_{d,k}$  as correlated Rayleigh fading represented as
\begin{align}
\label{model_h2lk}
&\mathbf{h}_{2,l,k}= \sqrt{\beta_{2,l,k}}\mathbf{R}_{IRS_{l,k}}^{1/2} \mathbf{z}_{l,k}, \\
\label{model_hdk}
&\mathbf{h}_{d,k}=\sqrt{\beta_{d,k}} \mathbf{R}_{BS_{k}}^{1/2} \mathbf{z}_{d,k},
\end{align}
where $\mathbf{R}_{IRS_{l,k}}$ is the correlation matrix of the channel from IRS $l$ to user $k$, $ \mathbf{R}_{BS_{k}}$  is the correlation matrix from the BS to user $k$, and $\mathbf{z}_{l,k} \sim \mathcal{CN}(0,\mathbf{I}_{N}) $ and $\mathbf{z}_{d,k} \sim \mathcal{CN}(0,\mathbf{I}_{M})$ describe the fast fading vectors of the IRS-user link and the BS-user link, respectively. The path loss factor is denoted as $\beta_{d,k}$ for the direct channel  and $\beta_{2,l,k}$ for the IRS-user $k$ link. 

Next we discuss the channel model for each $\mathbf{H}_{1,l}$. Many works on IRS-assisted systems, for example: \cite{los_channel,nadeem2020intelligent,los, maxmin, LIS_los}, assume the BS-IRS channel to be purely LoS. This assumption is quite practical and is supported in literature using two points. First, there will always exist a LoS path between the BS and IRS. As the BS tower is generally elevated high and the IRS is also envisioned to be integrated onto the walls of (high-rise) buildings, so both will have few obstacles around. Given the positions of BS and IRSs are fixed, a stable LoS channel between the BS and each IRS will exist and can be constructed at  the BS using directional (LoS AoD and AoA) information. Second, the path loss in NLoS paths is much larger than that in LoS path in the next generation systems, resulting in any NLoS paths constituting the BS-IRS channel to be much weaker. In fact it is noted that in mmWave systems, the typical value of Rician factor (ratio of energy in LoS to that in NLoS) is $20\rm{dB}$ and can be as large as $40$\rm{dB} in some cases \cite{los_channel}, which is sufficiently large to neglect any NLoS channel components.  Under these remarks, we assume that each BS-IRS $l$ channel is LoS. For uniform linear arrays at the BS and IRSs, the LoS channel gain between the $m^{th}$ BS antenna and the $n^{th}$ element in IRS $l$ is
\begin{align}
\label{model_H1}
&[\mathbf{H_{1,l}}]_{m,n}=  \sqrt{\beta_{1,l}}\exp\left(\frac{-j2\pi d_{m,n}}{\lambda_c}\right),
\end{align}
where $\beta_{1,l}$ the path loss factor for BS-IRS $l$ link and 
\begin{align}
    &d_{m,n} \approx d +  (n-1) \Delta_{IRS} \lambda_c \text{cos}(\theta_{{IRS}_{l,m}})\text{cos}( \phi_{{IRS}_{l,m}})\nonumber\\
    &- (m-1) \Delta_{BS} \lambda_c \text{cos}(\theta_{{BS}_{l,n}})\text{cos}( \phi_{{BS}_{l,n}})
\end{align}
where $d$ is the distance from the transmit antenna to the first IRS element,  $\Delta_{IRS} $ is the normalized IRS element separation, $\theta_{{IRS}_{l,m}}$ and $ \phi_{{IRS}_{l,m}}$ are the elevation and azimuth LoS AoD from $m^{th}$ BS antenna to IRS $l$, and $\theta_{{BS}_{l,n}}$ and $\phi_{{BS}_{l,n}}$ denote the elevation and azimuth AoA at the $n^{th}$ element of IRS $l$. 



\section{Channel Estimation Protocol}

This section presents the proposed CE protocol and derives analytical expressions of the MMSE estimates and NMSE.

\subsection{Proposed Protocol}
\label{protocol}

In this paper, we assume a time division duplexing (TDD) system where the downlink channels are estimated using the received uplink pilot symbols by exploiting channel reciprocity.   Since the IRSs are passive, the BS has to estimate all channels and determine optimal IRS configurations. 

We assign an uplink training phase of $\tau_C$ sec which is further broken down into $S$ sub-phases of length $\tau_S=\frac{\tau_C}{S}$ sec. Throughout the CE phase, users transmit mutually orthogonal pilot sequences of length $T_S$ symbols, with $T_S=\frac{\tau_S}{\tilde{\tau}}$ where $\tilde{\tau}$ is the duration of each symbol. The pilot sequence of user $k$ is denoted as $\mathbf{x}_{p,k}=[x_{p,k,1}, \dots, x_{p,k,T_S}]^T \in \mathbb{C}^{T_S \times 1}$, such that $\mathbf{x}_{p,k}^H\mathbf{x}_{p,l}=0$, for $k\neq l$, $k,l=1,\dots, K$ and $\mathbf{x}_{p,k}^H\mathbf{x}_{p,k}=P_C T_S \tilde{\tau}=P_C \tau_S$ Joules, where $P_C$  is the transmit power of each user. The IRS $l$ applies the reflect beamforming matrix $\boldsymbol{\Theta}_{l,s} = \text{diag}[\phi_{l,s,1}, \dots, \phi_{l,s,N}]^T \in \mathbb{C}^{N \times N}$ in sub-phase $s$, where $\phi_{l,s,n}=\alpha_{l,s,n} \exp(j\theta_{l,s,n})$. The received training signal, $\mathbf{Y}^{tr}_s\in \mathbb{C}^{M\times T_S}$ in sub-phase $s$ is then given as
\begin{align}
&\mathbf{Y}^{tr}_s=\sum_{k=1}^K (\mathbf{h}_{d,k}+\sum_{l=1}^{L}\mathbf{H}_{1,l} \boldsymbol{\Theta}_{l,s}\mathbf{h}_{2,l,k}) \mathbf{x}_{p,k}^H+ \mathbf{N}_s^{tr}
\end{align}
where  $\mathbf{N}^{tr}_s \in \mathbb{C}^{M\times T_S}$ is the matrix of noise vectors at the BS, with each column distributed independently as $\mathcal{CN}(\textbf{0},\sigma^2\textbf{I}_{M})$. After correlating the received training signal with the training sequence of user $k$, the BS obtains the observation vector, $\mathbf{r}^{tr}_{s,k}\in \mathbb{C}^{M\times 1}$, for user $k$ in sub-phase $s$ as
\begin{align}
\label{H_known1}
&\mathbf{r}^{tr}_{s,k}=(\mathbf{h}_{d,k}+\sum_{l=1}^{L}\mathbf{H}_{1,l}\boldsymbol{\Theta}_{l,s}\mathbf{h}_{2,l,k} )+\mathbf{n}^{tr}_{s,k},
\end{align}
which can be compactly written as
\begin{align}
\label{H_compact}
\mathbf{r}^{tr}_{s,k}&= \mathbf{h}_{d,k}+\mathbf{H}_{1}\boldsymbol{\Theta}_s\mathbf{h}_{2,k}
+\mathbf{n}_{s,k}^{tr} , \hspace{.02in} k=1,\dots, K,
\end{align}
where  $\mathbf{h}_{2,k}=[\mathbf{h}_{2,1,k}^T,\dots, \mathbf{h}_{2,L,k}^T]^T \in \mathbb{C}^{NL\times 1}$ is the concatenated vector of all $L$ IRS-user $k$ channels, $\boldsymbol{\Theta}_s =\text{diag}(\boldsymbol{\Theta}_{1,s} , \dots,\boldsymbol{\Theta}_{L,s} )  \in \mathbb{C}^{NL\times NL}$ is the reflection matrix of all the IRSs in sub-phase $s$, $\mathbf{H}_{1} = [\mathbf{H}_{1,l}, \dots, \mathbf{H}_{1,l}] \in \mathbb{C}^{M \times NL}$ is the concatenation of all $L$ BS-IRS channel matrices, and $\mathbf{n}_{s,k}^{tr} =\frac{\mathbf{N}^{tr}_s\mathbf{x}_{p,k}}{P_C \tau_S}\in \mathbb{C}^{M\times 1}$ is the noise.


Since we remarked earlier that $\mathbf{H}_{1,l} \in \mathbb{C}^{M \times N}$ can be perfectly known at the BS using locations of the IRSs, we will focus only on the estimation of $\mathbf{h}_{d,k}$s and $\mathbf{h}_{2,k}$s.  To this end, we define  $\bar{\mathbf{H}}_{1} = \text{diag}(\sqrt{M}\mathbf{I}_{M},\mathbf{h}_{1,1,1},  \dots, \mathbf{h}_{1,1,N}, \mathbf{h}_{1,2,1}  \dots, \mathbf{h}_{1,2,N},\dots, \mathbf{h}_{1,L,1},  \\ \dots, \mathbf{h}_{1,L, N} ) \in \mathbb{C}^{M(NL+1) \times NL+M}$ as the block diagonal matrix of $\mathbf{h}_{1,l,n}$s, where $\mathbf{h}_{1,l,n}$ is the $n$th column of  $\mathbf{H}_{1,l}$. Aggregating the observation vectors across the $S$ sub-phases as $\mathbf{r}^{tr}_{k}= [\mathbf{r}^{{tr}^T}_{1,k},\dots,\mathbf{r}^{{tr}^T}_{S,k}]^T \in \mathbb{C}^{SM \times 1}$, we obtain
\begin{align}
\label{Vtr}
\mathbf{r}^{tr}_{k}&= (\mathbf{V}_{tr} \otimes  \mathbf{I}_{M})\bar{\mathbf{H}}_{1}\mathbf{h}_{k}
+\mathbf{n}_{k}^{tr}, \hspace{.02in} k=1,\dots, K,
\end{align}
where,
\begin{align}
\label{Vtr_shape}
&\mathbf{V}_{tr}=\begin{bmatrix} 1 & \mathbf{v}_{1,1}^T & \dots & \mathbf{v}_{L,1}^T\\
	\vdots& 	\vdots & &\vdots\\ 
      1 & \mathbf{v}_{1,S}^T & \dots & \mathbf{v}_{L,S}^T
  \end{bmatrix} \in \mathbb{C}^{S\times (NL+1)},
	\end{align}
 $\mathbf{h}_{k}= [\frac{1}{\sqrt{M}}\mathbf{h}_{d,k}^T, \mathbf{h}_{2,k}^T]^T \in \mathbb{C}^{(NL+M) \times 1}$ cascades the direct and IRS-user channels and $\mathbf{{n}}_{k}^{{tr}^T}= [\mathbf{{n}}_{1,k}^{{tr}^T},\dots,\mathbf{{n}}_{S,k}^{{tr}^T}] ^T\in \mathbb{C}^{SM \times 1}$ is the concatenated noise across all sub-phases. Note  $\mathbf{v}_{l,s}=\text{diag}(\boldsymbol{\Theta}_{l,s}) $ is the beamforming vector for IRS $l$ in subphase $s$.

Define $\tilde{\mathbf{V}}_{tr}=(\mathbf{V}_{tr} \otimes  \mathbf{I}_{M})\bar{\mathbf{H}}_{1} \in \mathbb{C}^{SM \times NL+M}$. To guarantee the existence of left pseudo-inverse of $\tilde{\mathbf{V}}_{tr}$ (given as $ \tilde{\mathbf{V}}_{tr}^\dagger= (\tilde{\mathbf{V}}_{tr}^H \tilde{\mathbf{V}}_{tr})^{-1}\tilde{\mathbf{V}}_{tr}^H$), we must have $SM \geq NL+M$.  Rewriting the condition to accentuate only the parameter $S$, we obtain 
\begin{align}
&S \geq \frac{NL}{M}+1.
\end{align}
Thus the minimum number of CE sub-phases is actually reduced by a factor of $M$, as compared to the conventional protocols. For example: as compared to the protocol in \cite{MVUE} for a single IRS ($L=1$) system, which requires $N+1$ sub-phases to estimate the channels, we will require $N/M+1$ sub-phases, which is a substantial reduction given next generation systems use large numbers of BS antennas. Applying $ \tilde{\mathbf{V}}_{tr}^\dagger$ on \eqref{Vtr} results in 
\begin{align}
\label{Vtr2}
&\mathbf{\tilde{r}}^{tr}_k=\mathbf{h}_{k}+(\tilde{\mathbf{V}}_{tr}^H \tilde{\mathbf{V}}_{tr})^{-1}\tilde{\mathbf{V}}_{tr}^H \mathbf{{n}}^{tr}_{k}. 
	\end{align}
We denote by $\tilde{\mathbf{n}}_{k}^{tr} =(\tilde{\mathbf{V}}_{tr}^H \tilde{\mathbf{V}}_{tr})^{-1}\tilde{\mathbf{V}}_{tr}^H \mathbf{{n}}^{tr}_{k} \in \mathbb{C}^{(NL+M) \times 1}$ the measurement noise and by $\mathbf{C}_{\tilde{\mathbf{n}}_{k}^{tr}}= \mathbb{E}[\mathbf{\tilde{n}}_{k}^{tr}\mathbf{\tilde{n}}_{k}^{{tr}^H}] \in \mathbb{C}^{(NL+M)\times (NL+M)}$ the covariance matrix of this noise. 

\subsection{Noise Variance Minimization and MMSE Estimation }

 Next step is to design $\mathbf{V}_{tr}$ such that variance of the noise is minimized and the noise across the estimated channels is uncorrelated. 
 
 Recalling that $\mathbf{n}_{s,k}^{tr} =\frac{\mathbf{N}^{tr}_s\mathbf{x}_{p,k}}{P_C \tau_S} $ we obtain
\begin{align}
&\mathbf{C}_{\tilde{\mathbf{n}}^{tr}_k}=(\tilde{\mathbf{V}}_{tr}^H \tilde{\mathbf{V}}_{tr})^{-1}\tilde{\mathbf{V}}_{tr}^H \mathbb{E}\left[\mathbf{n}^{tr}_{k} \mathbf{n}^{tr^H}_{k}\right]{\tilde{\mathbf{V}}}_{tr}({\tilde{\mathbf{V}}}_{tr}^H {\tilde{\mathbf{V}}}_{tr})^{-1}, \\
\label{C_n}
&=\frac{\sigma^2 P_C \tau_S}{(P_C\tau_S)^2}({\tilde{\mathbf{V}}}_{tr}^H {\tilde{\mathbf{V}}}_{tr})^{-1}=\frac{\sigma^2}{P_C \tau_S}(\tilde{\mathbf{V}}_{tr}^H \tilde{\mathbf{V}}_{tr})^{-1},
\end{align}
where,
\begin{align}
\label{vtrhvtr}
&\tilde{\mathbf{V}}_{tr}^H \tilde{\mathbf{V}}_{tr}= \bar{\mathbf{H}}_{1}^H (\mathbf{V}_{tr}^H\mathbf{V}_{tr} \otimes  \mathbf{I}_{M})\bar{\mathbf{H}}_{1}.
\end{align}



We note that by construction $\bar{\mathbf{H}}_{1}^H  \bar{\mathbf{H}}_{1} =M\mathbf{\Sigma} \in \mathbb{C}^{NL+M \times NL+M}$, where $\mathbf{\Sigma} =\text{diag}(\mathbf{I}_M,\beta_{1,1}\mathbf{I}_N,...,\beta_{1,L}\mathbf{I}_N)$.
Therefore, to ensure uncorrelated channel estimates and equal noise variance in each estimate,  $ \mathbf{V}_{tr}$ must  have equally scaled orthogonal columns such that  $(\mathbf{V}_{tr}^H \mathbf{V}_{tr})^{-1}= \zeta \mathbf{I}_{NL+1}$.   Minimizing the noise variance is equivalent now to minimizing  $\zeta$, under the constraints: (1) $\mathbf{V}_{tr}$ has the structure in \eqref{Vtr_shape}, (2) $\alpha_{l,s,n}\in[0,1]$, (3) $\theta_{l,s,n} \in [0,2\pi]$, and (4) $(\mathbf{V}_{tr}^H \mathbf{V}_{tr})^{-1}=\zeta \mathbf{I}_{NL+1}$. 

The last constraint can be written as
\begin{align}
\label{trace}
&\zeta= \frac{NL+1}{tr(\mathbf{V}_{tr}^H \mathbf{V}_{tr})}= \frac{NL+1}{\sum_{s=1}^{S}\sum_{n=1}^{NL+1}|[\mathbf{V}_{tr}]_{s,n}|^2}.
\end{align}
Using the second constraint in \eqref{trace}, we see that 
\begin{align}
\label{lower_bound}
& \zeta \geq \frac{1}{S}.
\end{align}
 A solution that meets the lower bound in \eqref{lower_bound} with equality is the $NL+1$ leading columns of an $S \times S$ DFT matrix as expressed below \cite{nadeem2020intelligent, MVUE}.
 \begin{align}
\label{dft}
&[\mathbf{V}_{tr}]_{s,n}= w^{(n-1)(s-1)}, \hspace{.02in} n=1,\dots, NL+1, s=1,\dots, S,
\end{align}
where $w= e^{-j2\pi/S}$ is the primitive $S$th root of unity. This design of $\mathbf{V}^{tr}$ results in $(\mathbf{V}_{tr}^H\mathbf{V}_{tr})^{-1}=\frac{1}{S}\mathbf{I}_{NL+1}$ and therefore achieves the minimum noise variance  in \eqref{lower_bound} since $\zeta=\frac{1}{S}$. Using this we can simplify \eqref{vtrhvtr} as
\begin{align}
\label{VtrHvtr_simplifed}
&\tilde{\mathbf{V}}_{tr}^H \tilde{\mathbf{V}}_{tr}={S}\bar{\mathbf{H}}_{1}^H  \bar{\mathbf{H}}_{1} ={SM} \mathbf{\Sigma},
\end{align}
 and $(\tilde{\mathbf{V}}_{tr}^H\tilde{\mathbf{V}}_{tr})^{-1}= \frac{\mathbf{\Sigma}^{-1}}{SM}$.  

To derive the MMSE estimates, we refer back to \eqref{Vtr2} and simplify it with the optimal DFT design in \eqref{dft} to obtain
\begin{align}
\label{MMSE1}
&\tilde{\mathbf{r}}^{tr}_k=\mathbf{h}_{k}+\frac{1}{SM} \mathbf{\Sigma}^{-1}\tilde{\mathbf{V}}_{tr}^H \mathbf{{n}}^{tr}_{k} , \hspace{.02in} k=1,\dots, K,
	\end{align}


Recalling that $\mathbf{h}_{k} = [\frac{1}{\sqrt{M}}\mathbf{h}_{d,k}^T, \mathbf{h}_{2,k}^T]^T$, we can write $\tilde{\mathbf{r}}^{tr}_k = [ \tilde{\mathbf{r}}^{{tr}^T}_{0,k}, \tilde{\mathbf{r}}^{{tr}^T}_{1,k},\dots,\tilde{\mathbf{r}}^{{tr}^T}_{L,k}]^T\in \mathbb{C}^{(M+NL)\times 1}$, where $\tilde{\mathbf{r}}^{tr}_{0,k} \in \mathbb{C}^{M \times 1}$ is used to estimate the direct channel and $\tilde{\mathbf{r}}^{tr}_{l,k}\in \mathbb{C}^{N \times 1}$ is used to estimate the channel from IRS $l$ to user $k$.  

The relationship between   $\tilde{\mathbf{r}}_{0,k}$ and $ \mathbf{h}_{d,k}$ for user $k$ can be written using \eqref{MMSE1} and definition of $\tilde{\mathbf{V}}_{tr}$ as
\begin{align}
\label{MMSE1_direct2}
&\tilde{\mathbf{r}}^{tr}_{0,k}=\frac{1}{\sqrt{M}}\mathbf{h}_{d,k}+\frac{1}{SM} \mathbf{\Sigma}_0^{-1}\sqrt{M}\mathbf{I}_{M}(\mathbf{v}_1^{tr} \otimes \mathbf{I}_{M})^H \mathbf{{n}}^{tr}_{k} \nonumber\\
&=\frac{1}{\sqrt{M}}\mathbf{h}_{d,k}+\frac{1}{S\sqrt{M}} (\mathbf{v}_1^{tr} \otimes \mathbf{I}_{M})^H \mathbf{{n}}^{tr}_{k}.
\end{align}
where  $\sqrt{M}\mathbf{I}_{M}$  corresponds to the first $M\times M$ diagonal matrix of $\bar{\mathbf{H}}_1$, $ \mathbf{\Sigma}_0^{-1}=\mathbf{I}_M$ is the first $M\times M$ diagonal matrix of $ \mathbf{\Sigma}^{-1}$  and $\mathbf{v}_1^{tr}$ is the first $S \times 1$ column of  $\mathbf{V}_{tr}$. This observation vector can be scaled by a factor of $\sqrt{M}$ at the BS to obtain
\begin{align}
\label{MMSE1_direct}
&\tilde{\mathbf{r}}^{tr}_{0,k}=\mathbf{h}_{d,k}+\frac{1}{S} (\mathbf{v}_1^{tr} \otimes \mathbf{I}_{M})^H \mathbf{{n}}^{tr}_{k}.
\end{align}
Note that \eqref{MMSE1_direct} is actually the LS estimate of $\mathbf{h}_{d,k}$ which we denote as $\hat{\mathbf{h}}^{LS}_{d,k}$. The BS can compute the MMSE estimate of $\mathbf{h}_{d,k}$ as stated in the following lemma.
\begin{theorem}\label{L1}
The MMSE estimate $\hat{\mathbf{h}}_{d,k}$  of $\mathbf{h}_{d,k}$ is given as
\begin{align}
&\hat{\mathbf{h}}_{d,k}=\beta_{d,k}\mathbf{R}_{BS_k} \left(\beta_{d,k}\mathbf{R}_{BS_k}+ \frac{\sigma^2\mathbf{I }_M}{S P_C \tau_S}  \right)^{-1} \tilde{\mathbf{r}}^{tr}_{0,k},
\end{align}
which is distributed as $\hat{\mathbf{h}}_{d,k} \sim \mathcal{CN}(\mathbf{0}, \mathbf{\Psi}_{d,k})$ where
\begin{align}
\mathbf{\Psi}_{d,k}=
 \beta^2_{d,k} \mathbf{R}_{BS_k}\left(\beta_{d,k}\mathbf{R}_{BS_k}+ \frac{\sigma^2\mathbf{I }_M}{S P_C \tau_S}  \right)^{-1} \mathbf{R}_{BS_k}^H.
\end{align}

\end{theorem}
\begin{IEEEproof}
The proof follows from noting that the MMSE estimate is given as as $\hat{\mathbf{h}}_{d,k}=\mathbf{W}\tilde{\mathbf{r}}^{tr}_{0,k}$, where $\mathbf{W}=\mathbb{E}[\mathbf{h}_{d,k}\tilde{\mathbf{r}}^{{tr}^H}_{0,k}](\mathbb{E}[\tilde{\mathbf{r}}^{tr}_{0,k}\tilde{\mathbf{r}}^{{tr}^H}_{0,k}])^{-1}$. The expressions for $\mathbb{E}[\mathbf{h}_{d,k}\tilde{\mathbf{r}}^{{tr}^H}_{0,k}]$ and $\mathbb{E}[\tilde{\mathbf{r}}^{tr}_{0,k}\tilde{\mathbf{r}}^{{tr}^H}_{0,k}]$ can be derived by noting that $\mathbf{n}^{tr}_{k}$ and $\mathbf{h}_{d,k}$ are independent Gaussian vectors.
\end{IEEEproof}

The uncorrelated estimation error $\tilde{\mathbf{h}}_{d,k}= \hat{\mathbf{h}}_{d,k}-{\mathbf{h}}_{d,k}$ is also statistically independent of $\hat{\mathbf{h}}_{d,k}$ since both vectors are jointly Gaussian. Moreover, it is distributed as $\tilde{\mathbf{h}}_{d,k} \sim \mathcal{CN}(\mathbf{0}, \tilde{\mathbf{\Psi}}_{d,k})$, where $\tilde{\mathbf{\Psi}}_{d,k}=\beta_{d,k}\mathbf{R}_{BS_k}-\mathbf{\Psi}_{d,k}$.

Next we estimate $\mathbf{h}_{2,l,k} \in \mathbb{C}^{N \times 1}$ using the observation vector $\tilde{\mathbf{r}}^{tr}_{l,k}$. We extract $\bar{\mathbf{H}}_{1,l}= \text{diag}(\mathbf{h}_{1,l,1}, \dots, \mathbf{h}_{1,l,N})  \in \mathbb{C}^{MN \times N}$ from $\bar{\mathbf{H}}_1$ and note that $\bar{\mathbf{H}}_{1,l}^H \bar{\mathbf{H}}_{1,l}=M\mathbf{\Sigma}_l = \beta_{1,l}M\mathbf{I}_N $ for $ l=1, \dots, L$.  The relationship between $\tilde{\mathbf{r}}^{tr}_{l,k}\in \mathbb{C}^{N \times 1}$ and $\mathbf{h}_{2,l,k}$s can now be written using \eqref{MMSE1} as 
\begin{align}
\label{MMSE1_irs}
&\tilde{\mathbf{r}}^{tr}_{l,k}=\mathbf{h}_{2,l,k}+\frac{1}{SM}\mathbf{\Sigma}_l^{-1}\bar{\mathbf{H}}_{1,l}^H(\mathbf{V}_l^{tr} \otimes \mathbf{I}_{M})^H \mathbf{{n}}^{tr}_{k} \nonumber\\
&=\mathbf{h}_{2,l,k}+\frac{1}{SM\beta_{1,l}}\bar{\mathbf{H}}_{1,l}^H(\mathbf{V}_l^{tr} \otimes \mathbf{I}_{M})^H \mathbf{{n}}^{tr}_{k},
\end{align}
where $\mathbf{V}_l^{tr}=\mathbf{V}^{tr}(:,[N(l-1)+2:Nl+1]) \in \mathbb{C}^{S \times N}$ for $ l=1,\dots, L$.  Note that \eqref{MMSE1_irs} is actually the LS estimate of $\mathbf{h}_{2,l,k}$ which we denote as $\hat{\mathbf{h}}^{LS}_{2,l,k}$.

\begin{theorem} \label{L2} The MMSE estimate $\hat{\mathbf{h}}_{2,l,k}$ of $\mathbf{h}_{2,l,k}$ is given as
\begin{align}
\label{h2lkestimate_mmse}
\hat{\mathbf{h}}_{2,l,k}=\beta_{2,l,k}\mathbf{R}_{IRS_{l,k}}\left(\beta_{2,l,k}\mathbf{R}_{IRS_{l,k}}\hspace{-0.1 cm}+  \hspace{-0.1 cm}\frac{\sigma^2 \mathbf{I}_N}{\beta_{1,l}SMP_C\tau_S}\right)^{-1}\hspace{-0.35 cm}\tilde{\mathbf{r}}_{l,k},
\end{align}
for $ l=1,\dots, L$, $k=1,\dots, K$. It is distributed as $\hat{\mathbf{h}}_{2,l,k} \sim \mathcal{CN}(\mathbf{0}, \mathbf{\Psi}_{2,l,k})$, where 
\begin{align}
\mathbf{\Psi}_{2,l,k}=
 \beta^2_{2,l,k} \mathbf{R}_{IRS_k}\hspace{-0.15 cm}\left(\beta_{2,l,k}\mathbf{R}_{IRS_k}\hspace{-0.1 cm} +\hspace{-0.1 cm} \frac{\sigma^2 \mathbf{I}_N}{\beta_{1,l}SMP_C \tau_S}  \right)^{-1} \hspace{-0.35 cm}\mathbf{R}_{IRS_k}^H.
\end{align}
\end{theorem}
\begin{IEEEproof}
The proof is similar to that of $\hat{\mathbf{h}}_{d,k}$.
\end{IEEEproof}


The uncorrelated estimation error $\tilde{\mathbf{h}}_{2,l,k}= {\mathbf{h}}_{2,l,k}-\hat{\mathbf{h}}_{2,l,k}$ is also statistically independent of $\hat{\mathbf{h}}_{2,l,k}$ since both vectors are jointly Gaussian. Moreover, it is distributed as $\tilde{\mathbf{h}}_{2,l,k} \sim \mathcal{CN}(\mathbf{0}, \tilde{\mathbf{\Psi}}_{2,l,k})$, where $\tilde{\mathbf{\Psi}}_{2,l,k}=\beta_{2,l,k}\mathbf{R}_{IRS_k}-\mathbf{\Psi}_{2,l,k}$.

In order to compute the MMSE estimates, the BS must have previous knowledge of the correlation matrices at the IRSs and the BS. However, it is well-known that the correlation matrices vary slowly and remain static over many coherence intervals. The BS can easily obtain the prior knowledge of these matrices  since possessing second-order channel statistics at the BS is pervasive in massive MIMO literature \cite{nadeem2020intelligent}.

\subsection{NMSE Analysis}

Here we analyze the NMSE  in the LS and MMSE estimates under the developed CE protocol. We define the NMSE  (e.g. for the direct channel)  as
\begin{align}
&\text{NMSE}(\hat{\mathbf{h}}_{d,k})=\frac{tr(\mathbb{E}[\tilde{\mathbf{h}}_{d,k} \tilde{\mathbf{h}}_{d,k}^H])}{tr(\mathbb{E}[\mathbf{h}_{d,k}\mathbf{h}_{d,k}^H])}=\frac{tr(\tilde{\boldsymbol{\Psi}}_{d,k})}{tr(\mathbb{E}[\mathbf{h}_{d,k}\mathbf{h}_{d,k}^H])}.
\end{align}
The NMSE in the LS estimates is independent of the structure of $\mathbf{R}_{IRS_{l,k}} $ and $\mathbf{R}_{BS_{k}}$. The NMSE in the MMSE estimates does depend on these matrices, but we assume them to be identity matrices for the purpose of this section to obtain closed-form expressions and yield insights. The effect of correlation on NMSE is shown in the simulations later. 

First, the NMSE in the MMSE estimate of $\mathbf{h}_{d,k}$ is derived using the fact that $tr(\mathbb{E}[\mathbf{h}_{d,k}\mathbf{h}_{d,k}^H])= tr(\beta_{d,k} \mathbf{I}_M)=M \beta_{d,k}$ as
\begin{align}
&\text{NMSE}(\hat{\mathbf{h}}_{d,k})=\frac{\beta_{d,k}}{M\beta_{d,k}}\Big(M -\beta_{d,k}tr \Big(\beta_{d,k}\mathbf{I}_M+\frac{\sigma^2 \mathbf{I}_M}{S P_C \tau_S} \Big)^{-1}\Big)\nonumber \\
\label{sigma0}
&=\frac{1}{M\beta_{d,k}}\frac{M \beta_{d,k} \frac{\sigma^2}{S P_C \tau_S}}{\left(\beta_{d,k}+\frac{\sigma^2}{S P_C \tau_S}\right)}=\frac{\frac{\sigma^2}{S P_C \tau_S}}{\beta_{d,k}+\frac{\sigma^2}{S P_C \tau_S}}.
\end{align} 
The expression reveals that the NMSE increases to 1 as $\sigma^2$  grows large or $S$, $P_C$,  $\tau_S$ grow small (using L'Hopital's rule). Moreover by increasing $S$,  the quality of estimation can be improved but the training time will also increase. Similarly, the NMSE increases as $\beta_{d,k}$ grows small. 

The NMSE in the LS estimate of $\mathbf{h}_{d,k}$ in \eqref{MMSE1_direct} is obtained using $\mathbb{E}[\mathbf{n}_k^{tr}\mathbf{n}_k^{{tr}^H}]=\frac{\sigma^2 P_C \tau_S}{(P_C \tau_S)^2}\mathbf{I}_{MS}$ and $\mathbf{v}^{tr^H}_1 \mathbf{v}^{tr}_1=S$ as
\begin{align}
&\text{NMSE}(\hat{\mathbf{h}}_{d,k}^{LS})=\frac{\sigma^2 P_C \tau_Str\left((\mathbf{v}^{tr}_1 \otimes \mathbf{I}_M)^H (\mathbf{v}^{tr}_1 \otimes \mathbf{I}_M)\right)}{M\beta_{d,k}S^2 (P_C \tau_S)^2}\\
\label{exxpp3}
&=\frac{\sigma^2}{\beta_{d,k} S P_C \tau_S}.
\end{align}
We see that the NMSE in LS estimate will grow unboundedly if either $\sigma^2$ grows large, or the argument in the denominator grows too small. The difference  in the NMSE of MMSE and LS estimates can be straightforwardly calculated to be \\$\text{NMSE}(\hat{\mathbf{h}}^{LS}_{d,k})-\text{NMSE}(\hat{\mathbf{h}}_{d,k})$=
\begin{align}
&\frac{(\sigma^2)^2}{(\beta_{d,k} S P_C \tau_S)^2+\beta_{d,k} S P_C \tau_S\sigma^2}\geq 0,
\end{align}
for all $\beta_{d,k}$, $P_C$, $\tau_S$, and $S$ non-negative. Therefore the MMSE estimate of the direct channel will outperform the LS estimate. We can also observe the growth behavior of both NMSEs with respect to noise and see that the NMSE in MMSE estimate grows slower than that in the LS estimate, making MMSE robust even at low values of signal-to-noise ratio (SNR).

The NMSE in the estimates of $\mathbf{h}_{2,l,k}$s can be derived similarly by noting that from \eqref{model_h2lk} that under $\mathbf{R}_{IRS_{l,k}}=\mathbf{I}_N$, we have $tr(\mathbb{E}[\mathbf{h}_{2,l,k}\mathbf{h}_{2,l,k}^H])= tr(\beta_{2,l,k} \mathbf{I}_N)=N \beta_{2,l,k}$. Therefore $\text{NMSE}(\hat{\mathbf{h}}_{2,l,k})=\frac{1}{N\beta_{2,l,k}} tr(\tilde{\mathbf{\Psi}}_{2,l,k})$  and we get $\text{NMSE}(\hat{\mathbf{h}}_{2,l,k})=$
\begin{align}
&\frac{\beta_{2,l,k}}{N\beta_{2,l,k}}\Big(N -\beta_{2,l,k}tr \Big(\beta_{2,l,k}\mathbf{I}_N+\frac{\sigma^2\mathbf{I}_N}{\beta_{1,l}SM P_C \tau_S} \Big)^{-1}\Big), \nonumber \\
\label{mmse_h2lk_sigmal}
& = \frac{\frac{\sigma^2}{S MP_C \tau_S}}{\beta_{2,l,k}\beta_{1,l}+\frac{\sigma^2}{S MP_C \tau_S}}.
\end{align} 


Also, the NMSE in LS estimate in \eqref{MMSE1_irs} can be shown to be
\begin{align}
\label{lse_h2lk_sigmal}
&\text{NMSE}(\hat{\mathbf{h}}^{LS}_{2,l,k})=\frac{1}{\beta_{2,l,k}\beta_{1,l}}\frac{\sigma^2}{ SM P_C \tau_S}.
\end{align} 

The performance gap between the MMSE and LS estimate of $\mathbf{h}_{2,l,k}$ is $\text{NMSE}(\hat{\mathbf{h}}^{LS}_{2,l,k})-\text{NMSE}(\hat{\mathbf{h}}_{2,l,k})=$
\begin{align}
=\frac{(\sigma^2)^2}{(\beta_{2,l,k}\beta_{1,l} S MP_C \tau_S)^2+\beta_{2,l,k}\beta_{1,l}S MP_C \tau_S\sigma^2}\geq 0,
\end{align}
because $\beta_{2,l,k}$, $\beta_{1,l}$ and $M$ are  all non-negative. Therefore the MMSE estimate of $\mathbf{h}_{2,l,k}$ outperforms the LS estimate.

\section{Benchmark protocol}
As a benchmark (BM), we extend  the protocol in \cite{nadeem2020intelligent} for a single IRS to the distributed IRSs setup. The derivation is similar to \cite{nadeem2020intelligent}, and is repeated here for completeness. Under this protocol, we derive the estimates of the cascaded IRS-assisted channels, i.e. the columns of $\mathbf{H}_{0,l,k}=\mathbf{H}_{1,l}\text{diag}(\mathbf{h}_{2,l,k}) \in \mathbb{C}^{M \times N}$, without exploiting full knowledge of $\mathbf{H}_{1,l}$s. As a result, \eqref{H_compact} is written as
\begin{align}
\label{original_mmse}
&\mathbf{r}^{BM}_{s,k}=(\mathbf{h}_{d,k}+\mathbf{H}_{0,k} \mathbf{v}_{s})+\mathbf{n}^{tr}_{s,k}, \hspace{.02in} k=1,\dots, K, 
\end{align}
where $\mathbf{H}_{0,k}=\mathbf{H}_{1} \text{diag}(\mathbf{h}_{2,k}) \in \mathbb{C}^{M\times NL}$ is the cascaded matrix of all IRS-assisted channel vectors   and $\mathbf{v_s} = \text{diag}(\boldsymbol{\Theta}_s)\in \mathbb{C}^{NL\times 1}$ is the concatenated beamforming vector for all the IRSs in subphase $s$. Note that  $\mathbf{H}_{1}$, $\mathbf{h}_{2,k}$ and $\boldsymbol{\Theta}_s$ are defined after \eqref{H_compact}. We let $\bar{\mathbf{h}}_k=[\mathbf{h}_{d,k}^T, \mathbf{h}_{0,1,1,k}^T,\dots,\mathbf{h}_{0,1,N,k}^T, \mathbf{h}_{0,2,1,k}^T, \dots,  \mathbf{h}_{0,L,N,k}^T]^T \in \mathbb{C}^{M(NL+1)\times 1 }$, where $\mathbf{h}_{0,l,n,k}$ is the $n^{th}$ column of $\mathbf{H}_{0,l,k}$, and collect the observation vectors across $S$ subphases to obtain 
\begin{align}
\label{vtr_original}
&\mathbf{r}^{BM}_{k}=(\mathbf{V}_{tr} \otimes \mathbf{I}_{M})\bar{\mathbf{h}}_{k}+\mathbf{n}^{tr}_{k}, \hspace{.02in} k=1,\dots, K,
\end{align}
where $\mathbf{V}_{tr}$ is given as  \eqref{Vtr_shape} and $\mathbf{n}^{tr}_{k}$ is the same as in \eqref{Vtr}. To ensure that the left pseudo-inverse exists for $\mathbf{V}_{tr}$, we must have $S \geq NL+1$. Compared with the proposed protocol's condition of $S \geq \frac{NL}{M}+1$, the benchmark would impose almost an $M$ times larger training overhead which will compromise the downlink performance of the system. 

With optimal design for $\mathbf{V}_{tr}$ found to be the DFT matrix, we can perform the pseudo-inverse operation on \eqref{vtr_original} and get
\begin{align}
\label{vtr_original2}
&\tilde{\mathbf{r}}^{BM}_{k}=\bar{\mathbf{h}}_{k}+\frac{1}{S}(\mathbf{V}_{tr} \otimes \mathbf{I}_{M})^H\mathbf{n}^{tr}_{k}.
\end{align}

Write  $\tilde{\mathbf{r}}^{BM}_{k}=[\tilde{\mathbf{r}}^{BM^T}_{0,k}, \tilde{\mathbf{r}}^{BM^T}_{1,1,k},\dots, \tilde{\mathbf{r}}^{BM^T}_{1,N,k}, \dots, \tilde{\mathbf{r}}^{BM^T}_{L,1,k}, \dots, \\\tilde{\mathbf{r}}^{BM^T}_{L,N,k}]^T \in \mathbb{C}^{M(NL+1)\times 1 }$, we can obtain the MMSE estimates of  $\mathbf{h}_{d,k}$ and $\mathbf{h}_{0,l,n,k}$ as follows



\begin{align}
\label{h_d_est}
&\hat{\mathbf{h}}^{BM}_{d,k}=\beta_{d,k} \mathbf{R}_{BS_k}\mathbf{Q}_{d,k} \tilde{\mathbf{r}}^{BM}_{0,k},
\end{align}
where $\mathbf{Q}_{d,k}=\left(\beta_{d,k} \mathbf{R}_{BS_k}+\frac{\sigma^2}{S P_C \tau_S} \mathbf{I}_M \right)^{-1}$, and
\begin{align}
\label{h_irs_original}
&\hat{\mathbf{h}}^{BM}_{0,l,n,k}= r_{l,n,k} \beta_{l,k} \mathbf{h}_{1,l,n} \mathbf{h}^H_{1,l,n} \mathbf{Q}_{l,n,k}\tilde{\mathbf{r}}^{BM}_{l,n,k}, 
\end{align}
for $ n=1,\dots, N$, $l=1,\dots, L$, $k=1,\dots, K$, where $\mathbf{Q}_{l,n,k}=\left(r_{l,n,k} \beta_{l,k} \mathbf{h}_{1,l,n} \mathbf{h}_{1,l,n}^H  +\frac{\sigma^2}{SP_C \tau_S}\mathbf{I}_M\right)^{-1}$, $\beta_{l,k}=\beta_{1,l} \beta_{2,l,k}$, $r_{l,n,k}$ is the $(n,n)^{th}$ entry of the matrix $\mathbf{R}_{IRS_{l,k}}$ and $\mathbf{h}_{1,l,n}$ is the $n^{th}$ column of $\beta_{1,l}^{-1/2}\mathbf{H}_{1,l}$. 

Note that with  $\hat{\mathbf{H}}^{BM}_{0,l,k}=[\hat{\mathbf{h}}^{BM}_{0,l,1,k},\dots, \hat{\mathbf{h}}^{BM}_{0,l,N,k}] \in \mathbb{C}^{M\times N}$, the expression for $\hat{\mathbf{h}}^{BM}_{2,l,k}$ can be recovered under the benchmark protocol through the relationship $\hat{\mathbf{H}}^{BM}_{0,l,k}=\mathbf{H}_{1,l} \text{diag}(\hat{\mathbf{h}}^{BM}_{2,l,k})$, when $\mathbf{H}_{1,l} $ is known apriori. 


\begin{figure*}[t]
\centering
\subfloat[NMSE vs noise]{%
  \includegraphics[width=6cm]{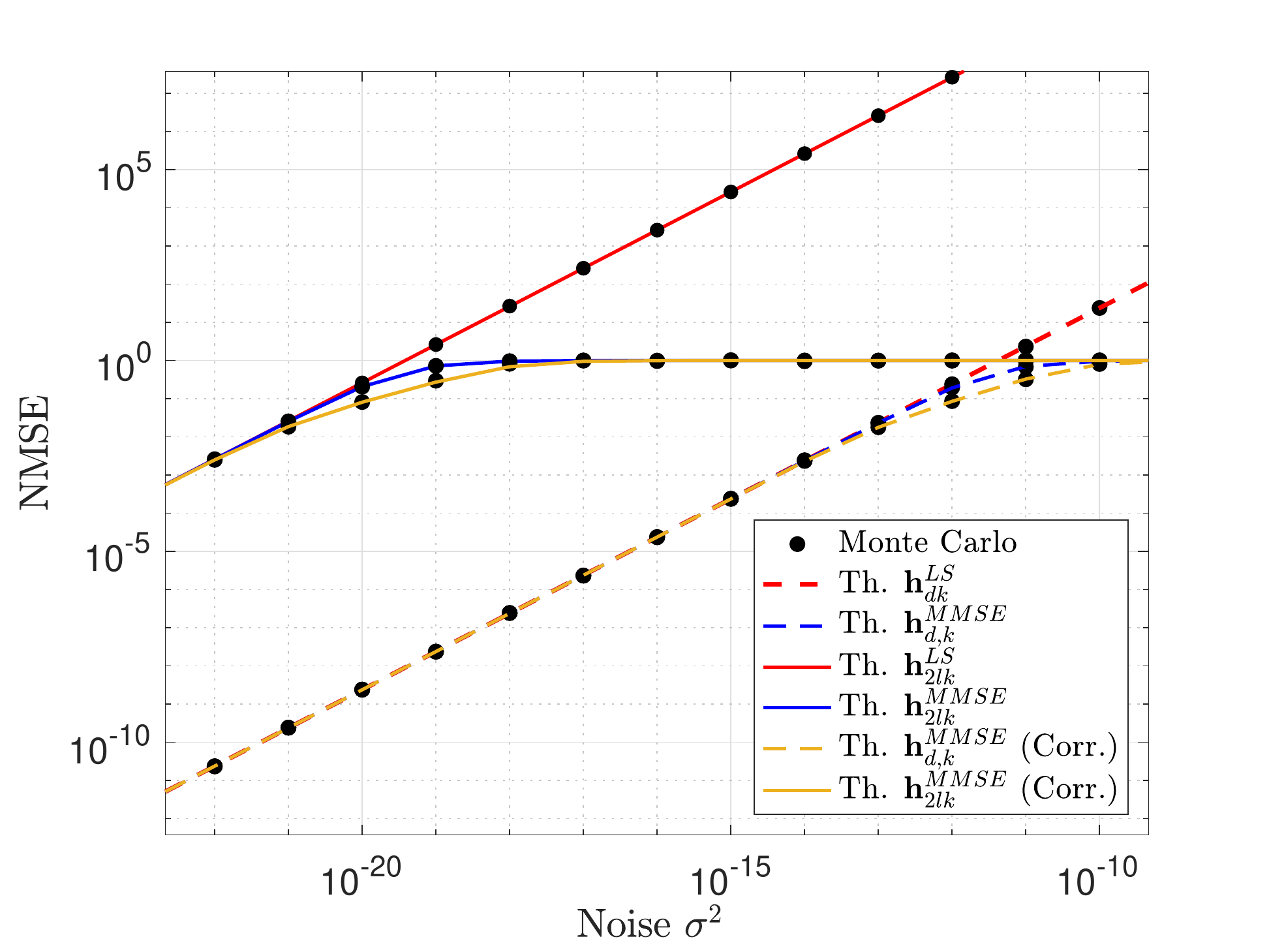}%
  \label{noiseNMSE}%
}
\subfloat[Direct Channel]{%
  \includegraphics[width=6cm]{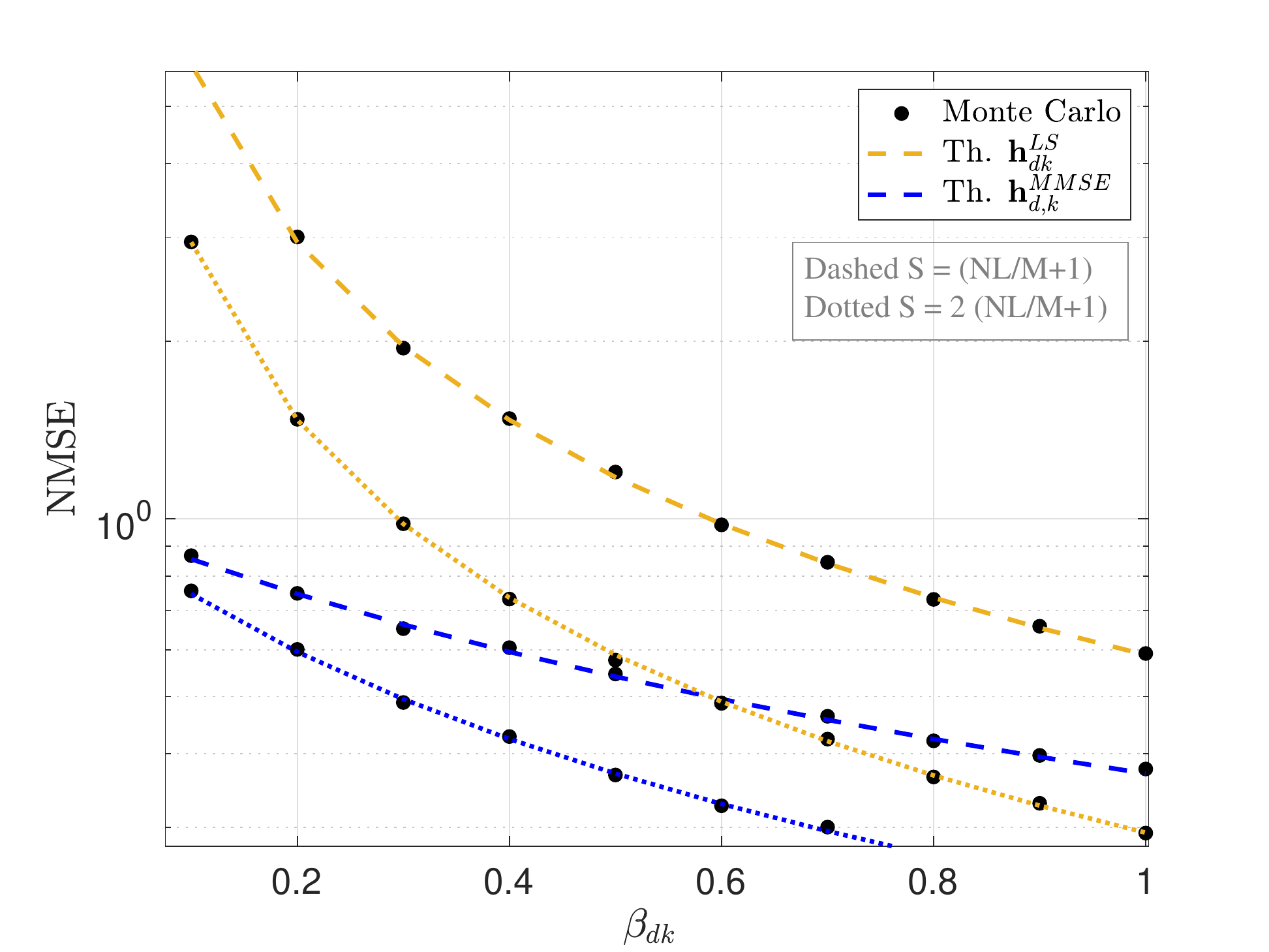}%
   \label{betadk}%
  }
\subfloat[IRS channel]{%
  \includegraphics[width=6cm]{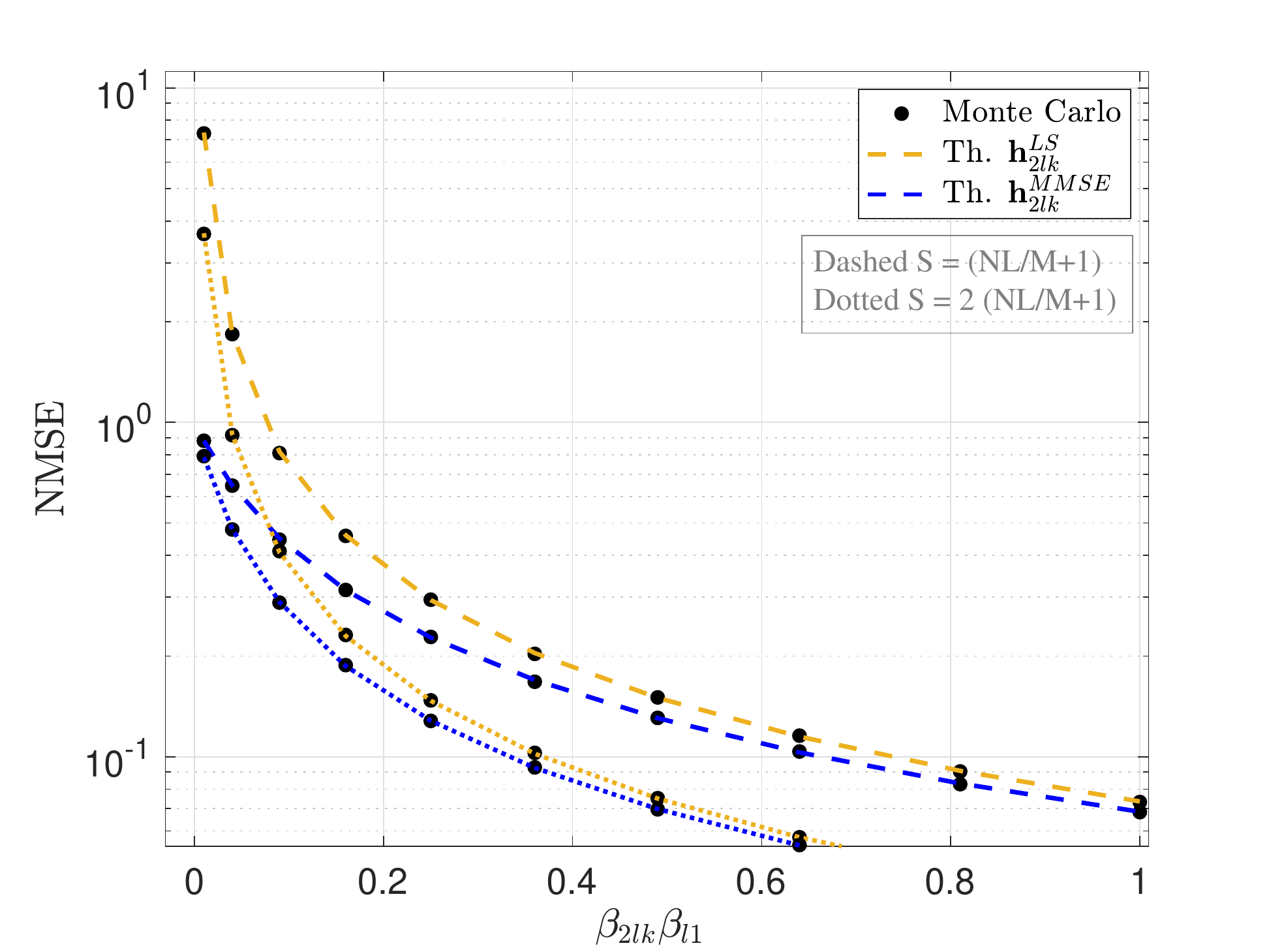}%
 \label{b2lk}%
}
\caption{NMSE vs noise and path loss factor for direct and IRS assisted links}
\end{figure*}

\section{Simulations and Discussion}
\label{Discussion}
In the simulation, we set $N=32$, $L=4$, $M=8$ and $\tau_S=50 \mu s$. Denoting $(x,y)$ metres (m) as the coordinates, the BS is placed at (0,0)m and the four IRSs are placed  at  (0,100), (100,0), (-100,0) and (0,-100)m. The path loss factors are computed as $\beta_{d,k}=\frac{10^{-2.8}}{d_{BS-user}^{3.67}}$, $\beta_{2,l,k}=\frac{10^{-2.8}}{d_{IRS l-user}^{3.67}} $ which are non-LoS (NLoS) path loss factors computed at 2.5 GHz carrier frequency for the 3GPP Urban Micro (UMi) scenario from TR36.814 (also found in Section V of \cite{nadeem2020intelligent}), and $\beta_{1,l}=\frac{10^{-2.6}}{d_{BS-IRS l}^{2.2}} $ is the LoS path loss factor, where $d_{(.)}$ stands for Euclidean distance for a particular link. The number of sub-phases is set as $S=\frac{NL}{M}+1=17$ for the proposed protocol. Fig. \ref{noiseNMSE} shows the NMSE curves for both MMSE and LS estimates of $\mathbf{h}_{d,k}$ and $\mathbf{h}_{2,l,k}$ (for the latter we plot $\frac{1}{L}\sum_{l=1}^L \text{NMSE}(\hat{\mathbf{h}}_{2,l,k})$). The Monte-Carlo simulated values match the theoretical analysis in Sec. III-C. As expected the MMSE estimates performs better than LS estimates especially in the high noise regime. Also, the NMSE in $\mathbf{h}_{2,l,k}$ is higher than that in $\mathbf{h}_{d,k}$ due to the multiplicative path loss factor $\beta_{1,l}\beta_{2,l,k}$ in \eqref{mmse_h2lk_sigmal} and \eqref{lse_h2lk_sigmal}. 

  The orange curves in the figure show the NMSE for the correlated scenario where $[\mathbf{R}_{IRS_{l,k}}]_{i,j}=\eta^{|i-j|}$ (similar definition for $\mathbf{R}_{BS_k}$) and $\eta=0.95$. The NMSE in the LS estimates is unaffected (not plotted for clarity of figures) while that in the MMSE estimates is actually less for the correlated case  \cite{nadeem2020intelligent}. 
	
	The NMSE vs the path loss factor is also plotted for the direct channel and IRS assisted channel in Fig. \ref{betadk} and Fig. \ref{b2lk} respectively, where it can be seen that the LS estimate performs worse than the MMSE estimates for all values of path loss factor. Increasing $S$ reduces the NMSE in both LS and MMSE estimates. However, increasing $S$ means increasing training overhead, which reduces the time left for downlink transmission.


In Fig. \ref{proposed}, we compare the proposed protocol which needs $S=\frac{NL}{M}+1$ sub-phases to the benchmark protocol which needs $S=NL+1$ sub-phases, in terms of NMSE and training time versus $L$. To ensure a fair comparison, both protocols are tested for the NMSE for the cascaded IRS-assisted channel.  The NMSE in the benchmark protocol is lower due to it using a larger $S$ for estimation, with the gap between the two NMSE curves becoming smaller with $L$. On the other hand, the time required for CE grows linearly with the number of IRSs in the benchmark protocol, which is detrimental for the overall downlink performance, while that for the proposed protocol grows with a much smaller slope. In fact, the gap in required training becomes significantly large with $L$ (or $N$). To better capture this trade-off, we tabulate in Tab. \ref{table1} a figure of merit defined as  $\frac{1}{NMSE \times \tau_C} (s^{-1})$, which can be interpreted as the measure of estimation accuracy per unit training time. Ideally, its value should be high. We observe that the proposed protocol actually performs better in terms of estimation accuracy per unit training time for all $L$. This makes the proposed protocol scalable to large-scale distributed deployment of IRSs.

\begin{figure}
\label{proposed}
\centering
  \includegraphics[scale=0.33]{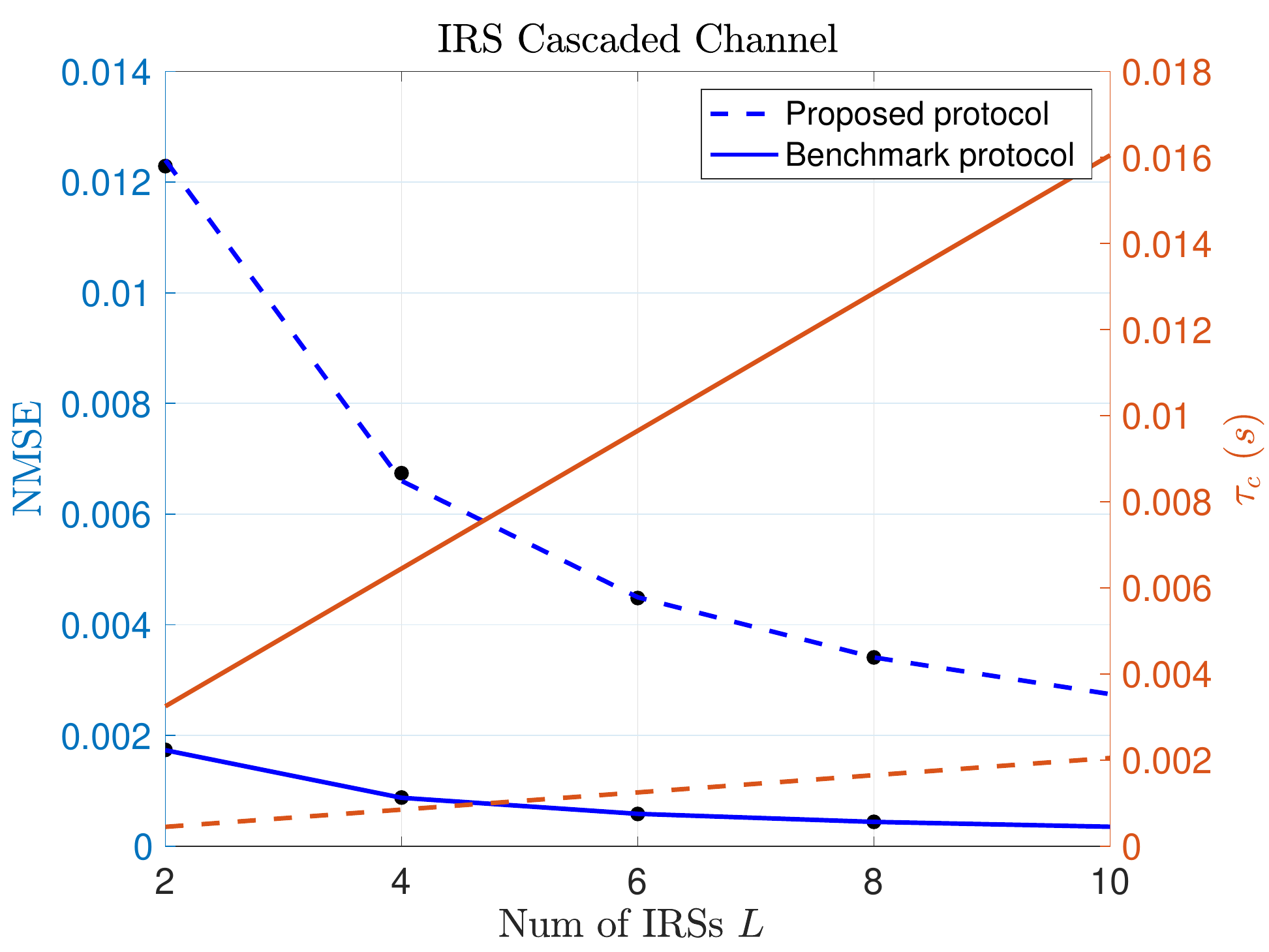}
    \caption{ Comparison of proposed and benchmark protocols.}
    \label{proposed}
\end{figure}

\begin{table}[]
\centering
\label{table1}
  \caption{ Estimation accuracy per unit time  $\frac{1}{NMSE \times \tau_C} (s^{-1})$}
\begin{tabular}{@{}lllll@{}}
\toprule
\textbf{L} & \textbf{2} & \textbf{6} & \textbf{8} & \textbf{10} \\ \midrule
Proposed  ($\times 10^5$) & 1.809      & 1.784       & 1.776        & 1.782        \\
Benchmark ($\times 10^5$)  & 1.767      & 1.777       & 1.766        & 1.769        \\ \bottomrule
\end{tabular}
\label{table1}
\end{table}

\section{Conclusion}
\label{conclusion}
This paper proposed a low-complexity CE protocol for the distributed IRSs-assisted MISO system and derived both LS and MMSE estimates of the BS-users direct channels and IRSs-users channels. An optimal solution for the phase shifts vectors at all IRSs is derived for the CE phase. 
By leveraging the LoS nature of the BS-IRSs channel matrices, the proposed protocol achieves a better estimation accuracy per unit training time as compared to the benchmark protocol.

\vspace{-.06in}
\bibliographystyle{IEEEtran}
\bibliography{bib.bib}

\end{document}